\documentclass{article}
\usepackage{arxiv}
\usepackage[utf8]{inputenc} 
\usepackage[T1]{fontenc}    
\usepackage{hyperref}       
\usepackage{url}            
\usepackage{booktabs}       
\usepackage{amsfonts}       
\usepackage{nicefrac}       
\usepackage{microtype}      
\usepackage{lipsum}
\usepackage{fancyhdr}       
\usepackage{graphicx}       
\graphicspath{{media/}}     

\usepackage{geometry}
\usepackage{pdflscape}
\usepackage{rotating}

\usepackage{pdfpages}
\usepackage{float}
\usepackage{siunitx}
\usepackage{tikz-network}
\usepackage{bm}
\RequirePackage{etex}
\usepackage{caption}
\usepackage{subcaption}
\usepackage{amsthm}
\newtheorem{theorem}{Theorem}[section]

\newtheorem{lemma}[theorem]{Lemma}
\newtheorem{proposition}[theorem]{Proposition}
\theoremstyle{definition}

\usepackage{setspace}
\title{Temporal Configuration Model:\\
Statistical Inference and Spreading Processes
}
\newcommand{\specificthanks}[1]{\@fnsymbol{#1}}

\author{
  Thien-Minh Le\footnote{Shared co-first authors.}
   \\
  Department of Mathematics \\
  University of Tennessee at Chattanooga\\
  Chattanooga, TN\\
  \texttt{thien-le@utc.edu} \\
   \And
   Hali Hambridge\footnotemark[\value{footnote}]\\
  Department of Biostatistics \\
  Harvard University \\
  Boston, MA\\
  \texttt{hhambridge@g.harvard.edu} \\
  \AND
  Jukka-Pekka Onnela \\
  Department of Biostatistics \\
  Harvard University \\
  Boston, MA \\
  \texttt{onnela@hsph.harvard.edu} \\
}
 \let\footnotesize\small
  
  \let \footnote \thanks

\begin{document}
\maketitle

\begin{abstract}
We introduce a family of parsimonious network models that are intended to generalize the configuration model to temporal settings. We present consistent estimators for the model parameters and perform numerical simulations to illustrate the properties of the estimators on finite samples. We also develop analytical solutions for basic and effective reproductive numbers for the early stage of discrete-time SIR spreading process. We apply three distinct temporal configuration models to empirical student proximity networks and compare their performance.
\end{abstract}


\section{Introduction}
In its simplest form, a network is a static collection of nodes joined together by edges. Many systems of interest can be represented as networks, enabling the study of such systems and leading to new insights. \cite{newman2018networks,roland20review} Broadly speaking, networks capture the pattern of interactions between elements of a system, reducing the system to a basic topological structure that facilitates analysis. One of the most commonly studied types of networks are interpersonal networks, wherein the network encodes interactions between individuals.

While network data is highly sought after, high quality network data can be quite rare. Traditional surveys, wherein researchers ask individuals about their contacts, are plagued with challenges. With egocentric surveys, researchers ask individuals about their contacts. While these surveys scale well, they often result in small network fragments and can be subject to recall bias and sampling bias. With sociocentric surveys, researchers ask participants to identify contacts from a roster. These surveys tend to give more complete picture than egocentric surveys, but they scale poorly and require a priori information on whom an individual might come into contact with. \cite{delva2016connectdot,brea2018ego} Both of these data collection methods are also quite burdensome and tend to provide very subjective network data. Furthermore, they typically result in coarse data; that is, researchers are only able to identify sets of contacts, but lack information on the proximity, frequency, and duration of their interactions, as well as how these interactions change over time.

These data collection challenges often lead researchers to treat networks as static, despite nearly all networks evolving over time. The use of static networks can be particularly problematic when studying dynamic phenomena that take place over networks, like spreading processes. Studies have shown that when the network evolves slowly relative to a spreading process, the dynamics can be approximated using a static network. When the network evolves very rapidly relative to the spreading process, the dynamics can be approximated well by a time-averaged version of the network. However, when the network and spreading process evolve at comparable time scales, the interplay between the two becomes important. \cite{holme2012temporal,holme15reviewTCM, holme21maptemporal, gage2020reviewTCM, mehdi23reviewTCM} 

Infectious diseases are a prime example of a spreading process, with the contagion propagating through the population via contact between an infected individual and a susceptible individual. Typically, infectious disease modeling is done via a fully-mixed model, which assumes that every individual in the population interacts with every other individual with equal probability or rate. However, in the real world, individuals typically have a small circle of contacts with whom they interact repeatedly. This was emphasized during the COVID-19 pandemic when many individuals practiced social distancing, choosing to interact with only a small group of people dubbed their ``social bubble'' or ``pod.'' \cite{hambridge2021,wu22temporalcovid} These repeated contacts, or persistent ties, between individuals are well-represented by a network structure and often lead to isolation of the contagion within a particular part of the network. However, as the network topology changes, paths to new, uninfected portions of the network can open up. As such, it is critical to account for both persistent ties and changes in network structure over time. An approach that captures both of these phenomena may offer the most realistic look at how diseases spread through populations.

Thankfully, recent technological advances have made high-fidelity individual-level proximity data more widely available, thereby allowing us to study empirical time-varying networks more precisely. Wearable sensors, like RFID tags and Bluetooth-enabled devices, such as smartphones, allow researchers to collect accurate, granular data in a passive manner. \cite{vanhems2013estimating,sapiezynski2019interaction,neto2021combining}  These technologies also allow investigators to capture proximity information with high precision, measuring the exact time and duration that two individuals are in close contact, along with their approximate distance from one another. This high-fidelity data allows us to quantify exposure, making it ideal for studying the spread of contagion. However, this begs the question: how much temporal granularity is needed to accurately capture spreading processes over networks? Very granular temporal resolution poses its own challenges, as it may capture unimportant interactions and erode privacy-protecting measures. Coarser collection can help preserve privacy and be less burdensome to store and analyze. However, if the network information is too coarse, we may miss key temporal dynamics, thereby inhibiting our ability to accurately model spreading processes over the network. 

Over the last decade, temporal networks have emerged as a fascinating problem that has received substantial research interest from a variety of disciplines. The majority of research focuses on developing various temporal models that can describe real-world data. Popular approaches include dynamic link modeling, activity-driven modeling, link-node memory modeling, and community dynamics modeling. \cite{holme13optimalstatic, perra12activitydriventemporal,vestergaard14linknodetemporal,tiago2017modelcommunity} However, research into temporal network model fitting and validation is still in early development. In  \cite{zhang2017randomdynamic}, the authors presented a generative model on random graphs where new edges form and dissolve at a constant rate, and proposed ways to estimate the rate for empirical data. Our study takes a novel approach to the temporal network, allowing more freedom in how edges are formed and dissolved, and focuses on the primary question of how much temporal network information is required to reliably retrieve the underlying generative model. Specifically, we propose a generative model for temporal networks which employs an edge persistence rate. The rate can be a simple constant, generated from a distribution, or some function of the empirical data. We provide consistent estimators for the model parameters. Interestingly, we found that constructing an estimator using as much temporal network information as possible does not necessarily result in a superior estimator. We also show how to fit the proposed models to an empirical data set with promising results. Our findings will provide practitioners with valuable insights to enable reliable estimations while balancing the cost, storage, and privacy concerns that come with collecting dynamic network data. 

\section{Temporal Network Model}
\subsection{Model Specification}

The configuration model (CM) is a widely used network model as it balances both realism and simplicity. \cite{newman2018networks,holme2012temporal,Bailey18CM}
Unlike many other network models, the configuration model incorporates arbitrary degree distributions. The exact degree of each individual node is fixed, which in turn fixes the total number of edges. To construct a network realization from a configuration model, one first specifies the total number of nodes in the network, denoted $N$. Then, for a given node $i$, one specifies the degree of that node, $k_{i}$, repeating this process for all nodes in the network. Each node is then given ``stubs'' equal to its degree, where a stub is simply an edge that is only connected to the node in question with its remaining end free. Pairs of stubs are then selected uniformly at random and connected to form edges until no stubs remain. The configuration model has several attractive properties. First, each possible matching of stubs is generated with equal probability. Specifically, the probability of nodes $i$ and $j$ being connected is given by $\frac{k_{i}k_{j}}{2m-1}$, where $m$ is the total number of edges. Because any stub is equally likely to be connected to any other, many of the properties of the configuration model can be solved exactly. While the configuration model does allow for multi-edges and self-edges, their probability tends to zero as $N$ approaches infinity. The configuration model is recognized as ``one of the most important theoretical models in the study of networks'' and many consider it one of the canonical network models. \cite{newman2018networks} When studying a new question or process, the configuration model is often the first model network scientists employ, making it an ideal foundation for our temporal network model.  \cite{newman2018networks, Bailey18CM} 

We present a time-varying version of the configuration model that we call the temporal configuration model (TCM). Like the configuration model, the TCM allows for an arbitrary degree distribution. Additionally, it encodes persistent ties, a key motivation for using and studying network models, by assigning each dyad a latent persistence probability.

To construct a TCM, one creates an initial network, $G_{0}$, via the standard configuration model algorithm. At each subsequent discrete time point $t$, a new network, $G_{t}$, is generated as follows. For each edge in $G_{t-1}$, a Bernoulli trial is conducted with success probability equal to that edge's latent persistence probability. That is, for an edge between nodes $i$ and $j$, we perform a Bernoulli trial with success probability $p_{ij}$. If the trial results in a success, the edge remains. If the trial is unsuccessful, the edge is broken, creating two stubs. After Bernoulli trials have been completed for all edges in $G_{t-1}$, stubs are matched uniformly at random with one another, forming edges in the new network, $G_{t}$. This process is repeated for each time step until one obtains the desired sequence of graphs, $G_{0}, G_{1}, \dots, G_{T}$.


The TCM can take on several intuitive forms by adjusting the latent edge-wise persistence probability. We begin by considering the two extremes. First, we can recover the standard configuration model simply by setting $p_{ij}=1$ for all $i$, $j$. This creates a static network such that $G_{t}=G_{0}$ for all times $t$. Second, if we set $p_{ij}=0$ for all $i$, $j$, we simply generate independent and identically distributed random draws of graphs from a configuration model ensemble. At each $t$, a configuration model with the specified number of nodes and degree sequence will be generated, but this instance will be independent of the previous network realizations. 

A more interesting scenario is obtained by setting $p_{ij}$ equal to some fixed $p \in (0,1)$ for all pairs $i$, $j$. That is, we specify a single or homogeneous persistence probability for all edges in the network. This parameter dictates how quickly the network changes over time. If $p$ is large, the network will change slowly as edges are highly likely to persist forward in time. If $p$ is small, the network will change rapidly. 

If we wanted the persistence probability to vary across the network, creating a more heterogeneous population, we can draw edge-level probabilities from some distribution for all node pairs. We could also imagine circumstances where it may make sense to encode some functional form for $p_{ij}$. For instance, edge-level persistence probabilities could be a function of node-level attributes if such information is available. We could envision a scenario where some individuals are more likely to stay within a particular social group while others may frequently move between groups, be it due to age, location, or some other factor or combination of factors. In this setting, we might construct $p_{ij}$ to be some function of node $i$'s attributes, $a_{i}$, and node $j$'s attributes, $a_{j}$. As such, every edge would have a distinct probability that would relate back to the two individuals in question. We could take this one step further and have edge-level persistence probabilities be a function of the relationship between the two nodes. For instance, familial relationships may have a higher persistence probability, while acquaintances may have a lower persistence probability.

\subsection{Inference}
\label{subsec:fittingTCM}

We focus on three of the intuitive forms outlined above, namely a single persistence probability for the entire network, edge-level persistence parameters drawn from a probability distribution, and edge-level persistence parameters that are product of node-level persistence parameters drawn from a probability distribution. In the first model, the entire network has a single persistence probability, $p$. We then consider the case where the persistence rate $p_{ij}$ of edge $(i,j)$ is generated from a given distribution $W$, i.e., $p_{ij}\sim W$, for all $i$, $j$. Finally, we consider the case where $p_{ij} = p_{i}p_{j}$, where $p_i \sim W$, for $i = 1,\cdots, N$ are independently drawn from a given distribution $W$. Notice that the first model is a special case of the second model as the distribution $W$ shrinks to a constant $p$. 

We observe that the future status of each edge after one step follows a Bernoulli distribution with mean equal to the persistence rate of that edge, and the status of each edge after two time steps follows a Bernoulli distribution with mean equal to the square of the persistence rate. The number of edges remaining after one or two time steps in the network is the sum of independent Bernoulli distributions. Thus, to estimate the first moment of the distribution, we can simply use the proportion of edges persisted one time step. To estimate the second moment of the distribution, we can use the proportion of edges that persisted two time steps. Next, we show that these estimators are consistent for each scenario.

\textit{Remark:} As long as $W$ is uniquely determined from its first $k$ moments, the proposed approach can specify the distribution of $W$.


\textbf{Model 1:} All edges of the network have a single persistence probability $p$ for some $0 \leq p \leq 1$. 

Let $X_{0} = x_0$ denote the number of edges in the initial graph, $G_{0}$. Notice that $x_0$ is a fixed constant for a given initial graph $G_0$. Let $X_{1}$ denote the number of edges that persist to time $t = 1$. That is, $X_{1} = \sum_{(i,j) \in G_0} \text{Bernoulli}(p)$. We can then use the ratio $Z_{1}=\frac{X_{1}}{X_{0}}$ to estimate the persistence probability $p$. Lemma \ref{consistentZ1:fixed} gives the properties of estimator $Z_1$. 

\begin{lemma}\label{consistentZ1:fixed}
$\frac{Z_1 -p}{\sqrt{\big( p(1-p)\big)/X_0}}$ converges to a standard normal distribution, as $X_0 \rightarrow \infty$.
\end{lemma}
\textit{Proof:} We have $Z_{1}=X_{1}/X_{0} = \sum_{(i,j) \in G_0} \text{Bernoulli}(p)/ X_{0} $. Since $\sum_{(i,j) \in G_0} \text{Bernoulli}(p)$ is the sum of independent Bernoulli random variables, applying the central limit theorem we have $\frac{Z_1 -p}{\sqrt{\big( p(1-p)\big)/X_0}}$ converges to a standard normal distribution as $X_0 \rightarrow \infty.$ $\hfill \square$

Thus, $Z_1$ is an unbiased and consistent estimator for $p$. 

Since the network continues to evolve over time, we can incorporate this information to refine our estimate of $p$. Following the same logic as in Lemma \ref{consistentZ1:fixed}, we can show that the ratio of edges remaining after each time step is also an unbiased and consistent estimator of $p$. To obtain a more precise estimator for the persistence probability $p$, we can use the average of ratios over all time steps, denoted $\bar{Z}$. 

Before discussing the proposed estimator $\bar{Z}$, we first walk through the evolution of the temporal network and set up the necessary notation. During the first time step, the broken edges of the original network $G_0$ form stubs and are rematched to create new edges. Let $Y_{1}$ denote the number of newly formed edges (excluding self-loops and multi-edges) at $t=1$. For large networks, the number of self-loops and multi-edges is negligible relative to the total number of edges; thus, $Y_{1}=X_{0}-X_{1}+o(1)$, where $o(1)$ captures the number of self-loops and multi-edges. Let $X_{1}^{+}$ denote the total number of edges at $t = 1$ after adding the new edges to the remaining edges. That is, $X_{1}^{+}=X_{1}+Y_{1} = X_{0} + o(1)$. Then, at $t = 2$, the number of edges that persist is $X_{2}\sim $Binomial$(X_{1}^{+},p)$ and the total number of edges after the rematching is $X_{2}^{+}=X_{1}^{+}+o(1)$. Similarly, for time $t = T$, $X_{T}$ is the total number of edges persisting from $X_{T-1}^{+}$ and $X_{T}^{+}$ is the number of edges after the rematching.

Let $Z_{t}$ denote the fraction of edges that persist to time $t$, i.e., $Z_{t}=\frac{X_{t}}{X_{t-1}^{+}}$, for $t = 1,\cdots,T$.  Then the proposed estimator $\bar{Z}$ is defined as 
$\bar{Z}=\frac{1}{T}\left(Z_{1}+Z_{2}+\cdots+Z_{T}\right).$ Next, we present some important properties of the proposed estimator $\bar{Z}$.

\begin{theorem}
\label{consistentZbar:fixed}
The estimator $\bar{Z}$ is an unbiased and consistent estimator of the persistence probability $p$. In addition, when $T = o(X_0)$ and $T\rightarrow\infty$, $\bar{Z}$ converges to $p$ at the rate of  $O_p\left(\sqrt{\frac{1}{X_0}} (\frac{1}{T})^{1/2-\delta}\right)$, for some small $\delta > 0$.
\end{theorem}
\textit{Proof:} It follows from Lemma \ref{consistentZ1:fixed} that each estimator $Z_k$ is an unbiased and consistent estimator for the persistence probability $p$, for all $k = 1,\cdots,T$. Therefore, 
 $\bar{Z}
    =\frac{1}{T}\, \sum_{i=1}^{T}Z_{i}$ is also an unbiased and consistent estimator of $p$.

Denote $S_{T}=\sum_{i=1}^{T}Z_{i}$ and $\gamma_{st}=E\left[(Z_{s}-p)(Z_{t}-p)\right]$ for all $s,\,t\in\{1,\dots,T\}$. We now evaluate the variance of $S_{T}$.  

\begin{align*}
    \text{Var}(S_{T})&=E\left[(S_{T}-pT)^{2}\right]  =E\left[\left(\sum_{i=1}^{T}Z_{i}-pT\right)^{2}\right] =\sum_{s=1}^{T}\sum_{t=1}^{T}E\left[(Z_{s}-p)(Z_{t}-p)\right]= \sum_{s=1}^{T}\sum_{t=1}^{T}\gamma_{st}.
\end{align*}

For any $s>t$, we have
\begin{align*}
    \gamma_{st}&=E\left[(Z_{s}-p)(Z_{t}-p)\right] = E(Z_{s}Z_{t})-p\,E(Z_{s})-p\,E(Z_{t})+p^{2} =E(Z_{s}Z_{t})-p^{2}-p^{2}+p^{2}\\
    &=E\left(\frac{X_{s}}{X_{s-1}^{+}}\,\frac{X_{t}}{X_{t-1}^{+}}\right)-p^{2} =E\left[E\left(\frac{X_{s}}{X_{s-1}^{+}}\,\frac{X_{t}}{X_{t-1}^{+}}\Big|
    X_{s-1}^{+}\right)\right]-p^{2} \\
    &=E\left[E\left(\frac{X_{s}}{X_{s-1}^{+}}\Big| X_{s-1}^{+}\right)\right]E\left(\frac{X_{t}}{X_{t-1}^{+}}\right)-p^{2} \ \  \ \text{by independence of $\frac{X_{t}}{X_{t-1}^{+}}$ and $E\left(\frac{X_{s}}{X_{s-1}^{+}}\Big| X_{s-1}^{+}\right)$}\\
    &=p\,E\left(\frac{X_{t}}{X_{t-1}^{+}}\right)-p^{2}=p^{2}-p^{2} = 0.
\end{align*}

\noindent Thus, $\gamma_{st}=0$ for any $s>t$. Similarly, $\gamma_{st}=0$ for any $s<t$. Therefore, $\text{Var}(S_{T}) =  \sum_{s=1}^{T}\sum_{t=1}^{T}\gamma_{st} = \sum_{s=1}^{T}\gamma_{ss}$. 

For $T = o(X_0)$, we have $X_{s-1}^{+} = X_0 + o(1)$, for all $s \in 1,\cdots, T$. Therefore,
\begin{align*}
\gamma_{ss} &=E\big( (Z_{s}-p)(Z_{s}-p)\big) =E(Z_{s}^{2})-p^2 =E\left[\left(\frac{X_{s}}{X_{s-1}^{+}}\right)^{2}\right] - p^2\\
&=E\left[E\left(\left(\frac{X_{s}}{X_{s-1}^{+}}\right)^{2}\Big|X_{s-1}^{+}\right)\right] - p^2 = E\left[\left(\frac{1}{X_{s-1}^{+}}\right)^{2} E\left(X_{s}^{2}\Big|X_{s-1}^{+}\right)\right] - p^2 \\
    &= E\left[\left(\frac{1}{X_{s-1}^{+}}\right)^{2} \left[X_{s-1}^{+}\,p\,(1-p)+(p\,X_{s-1}^{+})^{2}\right]\right] - p^2 \\
    & = E\left[\frac{p\,(1-p)}{X_{s-1}^{+}}\right] \leq  \frac{C_1}{X_0}, \quad \text{for some} C_1 >0.
\end{align*}

  This gives us $\text{Var}(S_{T}) =  \sum_{s=1}^{T}\gamma_{ss} \leq T \frac{C_1}{X_0}$. Applying the Chebyshev's, for some $C_2>0$ and $0<\delta<1/2$, we have

\begin{align*}
P\left(\left|\bar{Z}-p\right|\geq C_2 \sqrt{\frac{1}{X_0}} \left(\frac{1}{T}\right)^{1/2-\delta}\right)&\leq\frac{E\left[(\bar{Z}-p)^{2}\right]}{C_2^2 \frac{1}{X_0} (\frac{1}{T})^{1-2\delta}}=\frac{E\left[(S_{T}-T\,p)^{2}\right]}{T^{2}C_2^2 \frac{1}{X_0} (\frac{1}{T})^{1-2\delta}}\\
    &\leq \frac{T  \frac{C_1}{X_0}}{T^{2}C_2^2 \frac{1}{X_0} (\frac{1}{T})^{1-2\delta}}= \frac{C_1}{C_2^2  T^{2\delta}}\rightarrow 0.
\end{align*}

Thus, $ \bar{Z}$ converges to $p$ at the rate of $O_p\left(\sqrt{\frac{1}{X_0}} (\frac{1}{T})^{1/2-\delta}\right)$. $\hfill  \square$

\textit{Remark}: From Theorem \ref{consistentZ1:fixed} and Lemma \ref{consistentZ1:fixed}, we see that by incorporating information about network evolution over time, the estimator $\bar{Z}$ converges to the persistence probability $p$ faster than $Z_1$ by a factor of $(\frac{1}{T})^{1/2-\delta}$ for some small positive $\delta$.

\textbf{Model 2:} Now, consider drawing $p_{ij}$ from a given distribution $W$, i.e., $p_{ij}\sim W$ for all $i$, $j$. The simplest way to structure this model is to set the persistence probability for edge $(i,j)$ as $p_{ij}$, fixing it over time. Alternatively, one could fix the persistence probability of an edge over a time window of $T_{0}$ time steps for some $T_{0}>1$. Under this model, edge persistence probabilities are regenerated from $W$ anew for each time window. This latter approach may be more realistic in some settings. 

We will show that in either case the ratio of edges remaining after the first time step from the original network $G_0$ and the ratio of edges remaining after the first two steps from the original network $G_0$ can serve as good estimators for the first and second moment of the distribution $W$, respectively. For simplicity, we only provide the proofs for the estimator of the first moment; the proof for the estimator of the second moment is similar. Recycling the notation from Model 1, we will use $Z_1 = \frac{X_1}{X_0}$ to estimate the first moment $E(W)$. Lemma \ref{consistentZ1:random} gives us the convergence property of the estimator $Z_1$. 

\begin{lemma}\label{consistentZ1:random}
$Z_1$ is an unbiased and consistent estimator of the first moment $E(W)$ of distribution $W$, as $X_0 \rightarrow \infty$. 
\end{lemma}
\textit{Proof:}  We first prove that, for any given combination of edge persistence probabilities drawn from $W$, $Z_1$ is an unbiased and consistent estimator for the average of the edge persistence probabilities. 
We have $Z_{1}=X_{1}/X_{0} = \sum_{(i,j) \in G_0} \text{Bernoulli}(p_{ij})/ X_{0} $, where $\text{Bernoulli}(p_{ij})$ are independent variables. For simplicity, let us re-enumerate the sequence of independent $\text{Bernoulli}(p_{ij})$ for all $(i,j) \in G_0$ and denote them as $U_1, U_2, \cdots, U_{X_0}$, where $U_i \sim \text{Bernoulli}(p_i)$ for $i=1,\cdots,X_0$. Then $Z_1 = \frac{1}{X_0} \sum_{i=1}^{X_0}U_i$. Denote $\mu_{X_0} = \frac{1}{X_0} \sum_{i=1}^{X_0}p_i$ and $\sigma_{X_0} = \frac{1}{X_0} \sqrt{\sum_{i=1}^{X_0}p_i(1-p_i)}$.

Using the Lyapunov central limit theorem for independent non-identical distributions, we see that $(Z_1 - \mu_{X_0})/\sigma_{X_0} $ will converge to a standard normal distribution if we can verify the Lyapunov condition (\ref{Lyapunov}) below:

\begin{equation}
\label{Lyapunov}
    \lim_{X_0 \rightarrow \infty}\frac{1}{\sigma_{X_0}^{2+\delta}} \sum_{i=1}^{X_0}  \left(\frac{1}{X_0}\right)^{2+\delta}E\left\vert U_i - p_i\right\vert^{2+\delta} \rightarrow 0, \text{for some} \ \delta > 0.
\end{equation}

Since $0 \leq p_i \leq 1$ for all $i$, $\sum_{i=1}^{X_0}p_i(1-p_i)$ is of the same order as $\sum_{i=1}^{X_0}p_i$. Therefore, $\sum_{i=1}^{X_0}p_i(1-p_i) \rightarrow \infty$ as $X_0 \rightarrow \infty$. In addition, we also have \begin{align*}
    E\vert U_i - p_i\vert^{2+\delta} &= p_i (1 - p_i)^{2+\delta} + (1-p_i)p_i^{2+\delta} \\ 
    &\leq p_i(1-p_i)^2 + (1-p_i)p_i^2 = p_i(1-p_i).
\end{align*}
Therefore, 
\begin{align*}
    \lim_{X_0 \rightarrow \infty}\frac{1}{\sigma_{X_0}^{2+\delta}} \sum_{i=1}^{X_0}  \left(\frac{1}{X_0}\right)^{2+\delta}E\left\vert U_i - p_i\right\vert^{2+\delta} &= \lim_{X_0 \rightarrow \infty}\frac{1}{\big( \sum_{i=1}^{X_0}p_i(1-p_i) \big) ^{1+\delta/2}} \sum_{i=1}^{X_0}E\vert U_i - p_i\vert^{2+\delta}\\
    &\leq \lim_{X_0 \rightarrow \infty}\frac{\sum_{i=1}^{X_0}p_i(1-p_i)}{\big( \sum_{i=1}^{X_0}p_i(1-p_i) \big) ^{1+\delta/2}} \\
    &= \lim_{X_0 \rightarrow \infty}\frac{1}{\big( \sum_{i=1}^{X_0}p_i(1-p_i) \big) ^{\delta/2}} \rightarrow 0.\\
\end{align*} 

So, $\frac{Z_1 - \mu_{X_0}}{\sigma_{X_0}}$ converges to a standard normal distribution, as $X_0 \rightarrow \infty$.

The central limit theorem for random observations generated from distribution $W$ gives us $\mu_{X_0} = E(W) + O_p\left(\sqrt{\frac{1}{X_0}}\right)$. Furthermore, we also have that $E(Z_1) = E( \frac{1}{X_0} \sum_{i=1}^{X_0}p_i) = E(W)$. Therefore, $Z_1$ is an unbiased and consistent estimator of the first moment $E(W)$ of distribution $W$ as $X_0 \rightarrow \infty$. $\hfill  \square$

\begin{figure}[H]
  \centering
  \makebox[0pt]{\includegraphics[width=\textwidth]{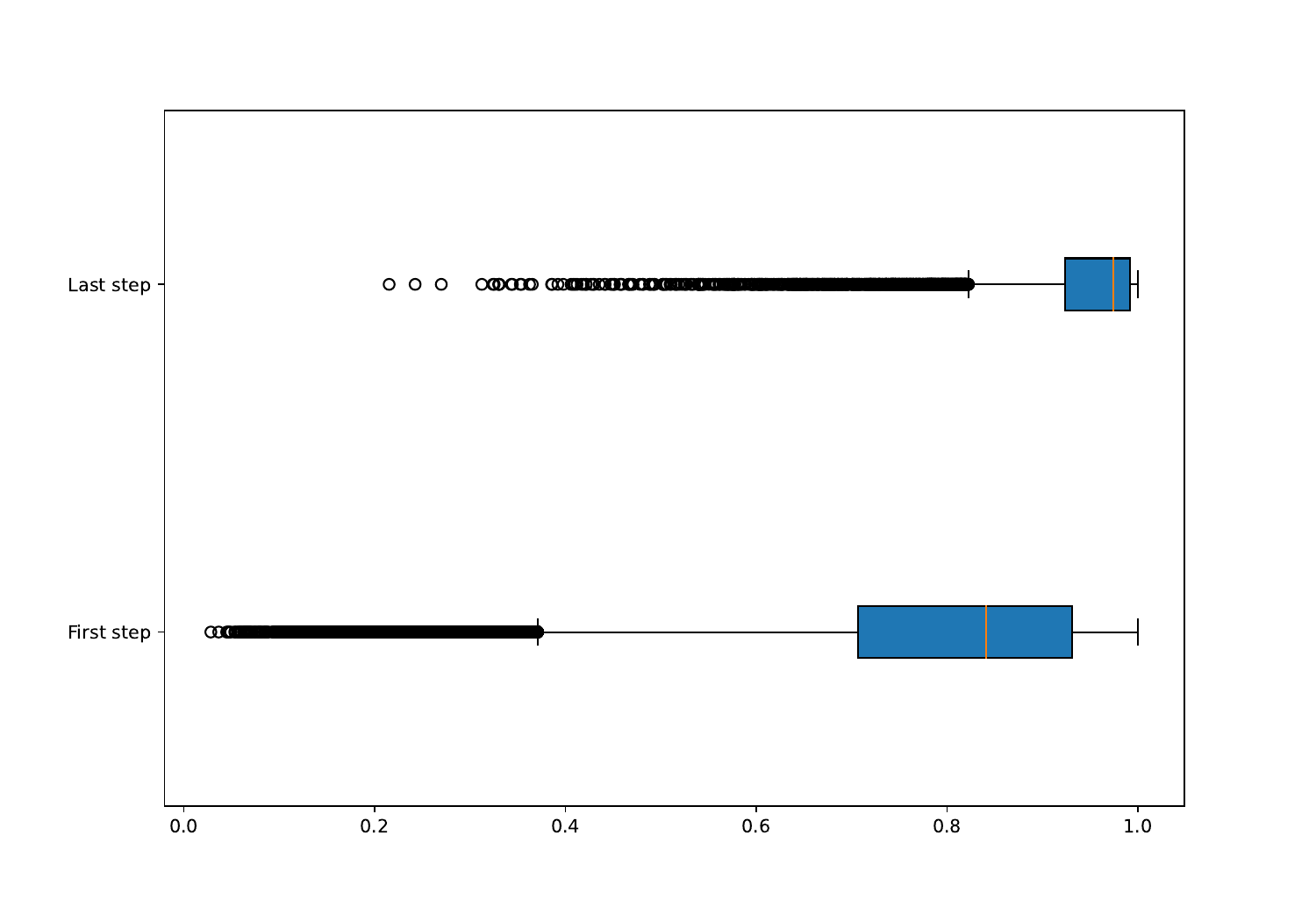}}
  \caption{\textbf{Box plots of edge persistence probabilities at the first step and the last step (step 100) for persistence edges of a network of size $1000$ over $100$ time steps. Edge persistent probabilities are generated from a $\text{Beta}(4,1)$ distribution.}}
  \label{fig:WeightRetained}
\end{figure}

Next, we consider how to utilize temporal network information to refine the estimator $Z_1$. Consider the case where we generate $p_{ij}$ from distribution $W$ and assign fixed persistence probabilities $p_{ij}$ for the $(i,j)$ dyad. In this case, $\bar{Z}$, constructed under Model 1, will not be a good estimator for the first moment $E(W)$ of $W$. As the network evolves, edges with higher persistence probabilities are more likely to be retained in the network. Therefore, over time, the persistence probabilities of edges remaining in the network will represent a shifted version of the original distribution $W$. As a result, using the estimator $\bar{Z}$ will result in a biased estimate of $E(W)$. Figure \ref{fig:WeightRetained} shows this phenomenon, where the initial network has $1000$ nodes and the edge persistence probabilities are generated from $W = \text{Beta}(4,1)$ for the first time steps. After $100$ time steps, edge persistence probabilities of edges retained in the network are shifted compared with the original distribution significantly.

Next, consider the variant of Model 2 where the edge persistence probabilities are generated anew from $W$ for each time window $T_0$ and kept fixed for each edge throughout each time window. More specifically, starting with $X_0$ edges from the set of edges $\bold{A}_0$ in the initial graph $G_0$ at time $t=0$, we generate persistence probabilities $p_{ij}^{(0)}, i,j = 1,\cdots, N$ from a distribution $W$ and assign the probability $p_{ij}^{(0)}$ to the edge $(i,j)$ if the edge appears at time $t = 0$ and persists to time $T_0-1$. At time step $T_0$, persistence probabilities $p_{ij}^{(1)}, i,j = 1,\cdots, N$ are generated anew from $W$ and assigned to the edge $(i,j)$ if the edge appears at time $t = T_0$ and persists to time $2T_0 - 1$. This process continues until we reach the desired number of time steps $T$. If additional information is available regarding the time window $T_0$, we can use information from the first $k$ time steps ($k < T_0$) from each window to obtain a better estimator for the first $k$ moments of distribution $W$.  

Denote the set of edges of $\bold{A}_0$ that persist to $t=1$ as $\bold{A}_1$. After the rematching process at $t=1$, the set of edges now becomes $\bold{A}_1^+$. Let $X_1$ denote the number of edges in $\bold{A}_1$ and $X_1^+$ the number of edges in $\bold{A}_1^+$ and so on for $t = 2,\cdots, T$. Let $\mathcal{F}_{k}^+$ denote the $\sigma-$algebra generated by all sets of edges up to time $k$, i.e., $\mathcal{F}_{k}^+ = \sigma \{\bold{A}_{1}^+, \dots, \bold{A}_{k}^+\}$. Finally, let $Z_k$ represent the fraction of edges remaining after one time step starting from each time window, i.e., $Z_1 = \frac{X_1}{X_0}, Z_2 = \frac{X_{T_0+1}}{X_{T_0}^+}, Z_3 = \frac{X_{2T_0+1}}{X_{2T_0}^+}, \cdots$. Then we can use $\bar{Z} = \sum_{k=1}^{m} Z_k/T$ to estimate the first moment $E(W)$, where $m = [T/T_0]$.

\begin{theorem}
\label{consistentZbar:random}
   If the periodic time interval $T_0$ is correctly specified, the proposed estimator $\bar{Z}$ is an unbiased and consistent estimator for $E(W)$. In addition, when $T = o(X_0)$ and $T\rightarrow\infty$, $\bar{Z}$ converges to $E(W)$ at the rate of  $O_p\left(\sqrt{\frac{1}{X_0}} \left(\frac{1}{m}\right)^{1/2-\delta}\right)$ for some small $\delta > 0, m = [T/T_0]$.
\end{theorem}

\textit{Proof:} It follows from Lemma \ref{consistentZ1:random} that each $Z_k$ is an unbiased and consistent estimator for the first moment $E(W)$, for $k = 1,\cdots,m$. Then $E(\bar{Z}) = E\left(\sum_{k=1}^{m} Z_k/m\right) = E(W).$ In other words, $\bar{Z}$ is also an unbiased and consistent estimator for $E(W)$, as $X_0, X_{T_0}, \cdots, X_{mT_0} \rightarrow \infty$.

To obtain the convergence rate of $\bar{Z}$, we first provide the upper bound for $E\big(\sum_{s=1}^{m} Z_s - m\,E(W)\big)^{2}$:

\begin{align*}
  E\left(\sum_{s=1}^m Z_s - m\,E(W)\right)^{2} &=  \text{Var}\left(\sum_{s=1}^{m} Z_s - m\,E(W)\right)  = \text{Var} \left( \sum_{s=1}^{m}  Z_s  \right) \\
  & = \sum_{s=1}^{m} \text{Var} (Z_s)  +  2 \sum_{s>t; s, t =1}^{m} \text{Cov} ( Z_s , Z_t ) .
\end{align*}

For any $s>t$, we have 
\begin{align*}
  \text{Cov} ( Z_s  , Z_t   ) 
  & =E\left(\frac{X_{(s-1)T_0+1}}{X_{(s-1)T_0}^{+}}\cdot \frac{X_{(t-1)T_0+1}}{X_{(t-1)T_0}^{+}}\right) -  \big( E(W) \big)^2    \\
    &=E\left[E\left(\frac{X_{(s-1)T_0+1}}{X_{(s-1)T_0}^{+}}\cdot \frac{X_{(t-1)T_0+1}}{X_{(t-1)T_0}^{+}}\Bigg|
    \mathcal{F}_{(s-1)T_0}^{+}\right)\right] -  \big( E(W) \big)^2   \\
     &=E\left[E\left(\frac{X_{(s-1)T_0+1}}{X_{(s-1)T_0}^{+}}\Bigg|
     \mathcal{F}_{(s-1)T_0}^{+}\right)\right] E\left( \frac{X_{(t-1)T_0+1}}{X_{(t-1)T_0}^{+}} \right) -  \big( E(W) \big)^2   \\
    &= E(W)   E(W)  -  \big( E(W) \big)^2   =  0.
\end{align*}

In addition, 
\begin{align*}
    \text{Var} ( Z_s )  &= E\left[\left(\frac{X_{(s-1)T_0+1}}{X_{(s-1)T_0}^{+}}\right)^2\right] -  \big( E(W) \big)^2  = 
    E\left[E\left(\left(\frac{X_{(s-1)T_0+1}}{X_{(s-1)T_0}^{+}}\right)^2 \Bigg| \mathcal{F}_{(s-1)T_0}^{+} \right)\right] -  \big( E(W) \big)^2\\
    &= E\left[ \left( \frac{1}{X_{(s-1)T_0}^{+}} \right)^2 E \left( X_{(s-1)T_0+1}^2 \Bigg| \mathcal{F}_{(s-1)T_0}^{+} \right) \right] -  \big( E(W) \big)^2.\\
\end{align*}

Since $X_{(s-1)T_0+1}$ is the summation of independent Bernoulli variables $\text{Bernoulli}\left(p_{ij}^{(s)}\right)$, where $(i,j)\in\bold{A}_{(s-1)T_0}^+$, we have 
\begin{align*}
    E \left( X_{(s-1)T_0+1}^2 \Big| \mathcal{F}_{(s-1)T_0}^{+} \right) &= \text{Var}\left(X_{(s-1)T_0+1} \vert \mathcal{F}_{(s-1)T_0}^{+}  \right) + \left[ E \left(X_{(s-1)T_0+1}\vert \mathcal{F}_{(s-1)T_0}^{+} \right) \right]^2 \\
    &= \sum_{(i,j) \in \bold{A}_{(s-1)T_0}^+ } p_{ij}^{(s)}\left(1-p_{ij}^{(s)}\right)   +   \left(\sum_{(i,j) \in \bold{A}_{(s-1)T_0}^+ } p_{ij}^{(s)}\right)^2 .
\end{align*}

For $T = o(X_0)$, we have $X_{(s-1)T_0}^{+} = X_0 + o(1)$, for all $s = 1, \cdots, m$. Therefore, 
\begin{align*}
    \text{Var} ( Z_s ) &=   E\left( \frac{ \sum_{(i,j) \in \bold{A}_{(s-1)T_0}^+ } p_{ij}^{(s)}\left(1-p_{ij}^{(s)}\right)}{ X_{(s-1)T_0}^{+ 2}} \right) +  E\left(\frac{\left(\sum_{(i,j) \in \bold{A}_{(s-1)T_{0}}^+ } p_{ij}^{(s)}\right)^2} {X_{(s-1)T_0}^{+ 2}}\right)   -  \big( E(W) \big)^2\\
    &\leq  E\left( \frac{ \sum_{(i,j) \in \bold{A}_{(s-1)T_0}^+ } p_{ij}^{(s)}}{ X_{(s-1)T_0}^{+ 2}} \right) + E\left(\frac{\text{Var}(W)}{X_{(s-1)T_0}^{+}}\right) \leq \frac{C}{X_0}, \, \text{for some}\, C >0.\\
\end{align*}

This gives us $E\big(\sum_{s=1}^{m} Z_s - m\,E(W)\big)^{2} \leq \frac{mC}{X_0}$. Using the same argument as in the proof of Theorem \ref{consistentZbar:fixed}, $\bar{Z}$ converges to $E(W)$ at the rate of  $O_p\left(\sqrt{\frac{1}{X_0}} (\frac{1}{m})^{1/2-\delta}\right)$, for some small $\delta > 0. \hfill \square$

\textbf{Model 3:} In Model 3, we consider the case where $p_{ij} = p_{i}p_{j}$, where $p_i, i = 1,\cdots, N$ are independently draw from a distribution $W$. As with Model 2, we consider two model variants. Since the persistence probability $p_{ij} = p_{i}  p_j$, $E(p_{ij}) = E(p_ip_j) = E(p_i) E(p_j) = E(W)^2$ and $E(p_{ij}^2) = E(p_i^2 p_j^2) = E(p_i^2) E(p_j^2) = E(W^2)^2$. Therefore, we can use the same estimation strategy as above to estimate moments of the distribution $W$. The estimator $\bar{Z}$ will only be beneficial for the estimation process if we can correctly specify the time window $T_0$. Using the same arguments as in Lemma \ref{consistentZ1:random} and Theorem \ref{consistentZbar:random}, we obtain the following: 

\begin{lemma}
\label{Z1consistent:productrandom}
    The proposed estimator $Z_1$ is an unbiased and consistent estimator for $E(W)^2$.
\end{lemma}

\textit{Proof:}
As in Lemma \ref{consistentZ1:random}, for any given combination of edge persistence probabilities generated from distribution $W$, the standardized version of $Z_1$ converges to a standard normal distribution. Applying the Central Limit Theorem to the degenerate U-statistics (Theorem 2.1 in  \cite{Alex2022}), we have  $\mu_{X_0} = \frac{1}{X_0} \sum_{(i,j)\in G_0} p_i p_j  =  E(W)^2 + O_p\left(\sqrt{\frac{1}{N}}\right)$. Furthermore, we also have that $E(Z_1) = \frac{1}{X_0} \sum_{(i,j)\in G_0} p_i p_j  =  E(W)^2$. Therefore, $Z_1$ is an unbiased and consistent estimator for $E(W)^2$, as $X_0 \rightarrow \infty$. 
$\hfill  \square$

\begin{theorem}
\label{consistentZbar:productrandom}
    If the periodic time interval $T_0$ is correctly specified, the proposed estimator $\bar{Z}$ is an unbiased and consistent estimator for $E(W)^2$. In addition, when $T = o(X_0)$ and $T\rightarrow\infty$, $\bar{Z}$ converges to $E(W)^2$ at the rate of  $O_p\left(\sqrt{\frac{1}{N}} (\frac{1}{m})^{1/2-\delta}\right)$ for some small $\delta > 0, m = [T/T_0]$.
\end{theorem}
\textit{Proof:} The proof follows by using the same arguments as in Theorem \ref{consistentZbar:random}.



 \section{Simulation Studies}
In this section, we perform some numerical simulations to illustrate the properties of proposed estimators on finite samples. We consider three different sets of simulation studies for each of the three models. The first set of simulations corresponds Model 1, where all edge persistence probabilities are fixed at $p = 0.8$. The second set of simulations examines Model 2, where edge persistence probabilities are generated from a $\text{Beta}(1,4)$ distribution, kept fixed for a time window of $T_0 = 2$, and then resampled at the start of a new window. More specifically, persistence probabilities for any given edge $(i,j)$ of the network at any two consecutive time steps $2(k-1)$ and $2(k-1) + 1$  are the same and drawn from $\text{Beta}(1,4)$, for $k = 1,\cdots, [T/2]$. Finally, we consider Model 3, where the persistence probability $p_{ij}$ of a given edge $(i,j)$ is the product of $p_i$ and $p_j$, where node-level persistence is drawn from $\text{Beta}(1,4)$, kept fixed for any two consecutive time steps $2(k-1)$ and $2(k-1) + 1$ and resampled at the start of a new window. To understand the effect of network size and number of time steps, we consider three different network sizes $N = 10$, $100$, and $1000$. For each $N$, we allow the network to evolve through $T = 30$ and $T = 100$ time steps. The original graph $G_0$ is first generated via the standard configuration model, where the degree sequence is generated from a Poisson distribution with mean $6$. The network then evolves through $T$ time steps with the persistence probabilities corresponding to each of the settings outlined above. 

We use the proposed estimators $Z_1$ and $\bar{Z}$ to estimate the persistence probability $p$ in the first set of simulations and the first moment of the underlying distribution in the second and third simulations. To estimate the second moment of the underlying distribution in the last two simulations, we use estimators $V_1$ and $\bar{V}$, where $V_1$ denotes the proportion of edges remaining after the first two time steps and $\bar{V}$ utilizes the average of $V_k$ when the time window $T_0$ is specified correctly. To evaluate the proposed estimators' accuracy, we compute the absolute relative bias and the standard deviation of each estimator based on $100$ replications. The absolute relative bias of estimators $Z_1$ and $\bar{Z}$ are defined as $\frac{1}{100}\sum\limits_{k=1}^{100} \Big\vert \frac{ Z_1^{(i)} - p }{p} \Big\vert$ and $\frac{1}{100}\sum\limits_{k=1}^{100} \Big\vert \frac{ \bar{Z}^{(i)} - p }{p} \Big\vert$, respectively, for the first set of simulations where, $Z_1^{(i)}$ and $\bar{Z}^{(i)}$ correspond to the $i$th replication. For the second and third simulation, the absolute relative biases are $\frac{1}{200} \left( \sum\limits_{k=1}^{100} \Big\vert \frac{ Z_1 - E(W) }{E(W)} \Big\vert + \sum\limits_{k=1}^{100} \Big\vert \frac{ V_1 - E(W^2) }{E(W^2)} \Big\vert \right) $ and $\frac{1}{200} \left( \sum\limits_{k=1}^{100} \Big\vert \frac{ \bar{Z} - E(W) }{E(W)} \Big\vert + \sum\limits_{k=1}^{100} \Big\vert \frac{ \bar{V} - E(W^2) }{E(W^2)} \Big\vert \right) $. We denote absolute relative bias and its standard deviation as AbsRelBias and SdAbsRelBias, respectively.

\begin{table}
\begin{center}
\captionof{table}{Absolute relative bias and standard deviations of the proposed estimators $Z_1$ and $\bar{Z}$ for Model 1 when edge persistence probability is fixed at $p = 0.8$.
\label{sim:fixedconsistent}}
\begin{tabular}{ccrcccc}
\hline
    &  &  \multicolumn{2}{c}{Estimator  $\bar{Z}$ } & \multicolumn{2}{c}{Estimator $Z_1$}\\
 $N$ & $T$  &  AbsRelBias & SdAbsRelBias & AbsRelBias & SdAbsRelBias \\ 
  \hline    
10 & 30 & 0.0022 & 0.0297 & 0.0097 & 0.1052 \\ 
  & 100 & 0.0017 & 0.0205 & 0.0023 & 0.1139 \\ 
   \cline{2-6}
  100 & 30 & 0.0007 & 0.0056 & 0.0002 & 0.0323 \\ 
  &  100 & 0.0005 & 0.0037 & 0.0035 & 0.0257 \\ 
   \cline{2-6}
  1000 & 30 & 0.0000 & 0.0017 & 0.0007 & 0.0093 \\ 
  &  100 & 0.0001 & 0.0008 & 0.0006 & 0.0095 \\
 \hline 
\end{tabular}
\end{center}
\end{table}

Table \ref{sim:fixedconsistent} shows that both $Z_1$ and $\bar{Z}$ are good estimators of $p$ with the estimator $\bar{Z}$ outperforming the estimator $Z_1$. The estimator $Z_1$ seems to be a reasonable estimator for $p$ when the network is greater than or equal to $N=100$, but does not perform well when the network size is $10$. $\bar{Z}$, on the other hand, is a good estimator of $p$ in all cases. These results suggest that $\bar{Z}$ is a more reliable estimator for $p$ when working with Model 1, where all edges have a fixed probability $p$. 

Tables \ref{sim:randomconsistent} and \ref{sim:productrandomconsistent} also demonstrate the consistency of the proposed estimators for large networks. As expected, with more information in hand, $\bar{Z}$ also outperforms to the estimator $Z_1$ as the number of time steps $T$ increases. For a small network size of $N=10$, $Z_1$ does a poor job while the estimator $\bar{Z}$ improves as the number of time steps increases from $T=30$ to $T=100$. As the network size increases, both estimators become more reliable and tend to concentrate at the underlying true value of the generating distribution.  

\begin{table}
\begin{center}
\captionof{table}{Absolute relative bias and standard deviations of the proposed estimators $Z_1$ and $\bar{Z}$ for Model 2 when edge persistence probabilities $p_{ij}$ are drawn from $\text{Beta}(1,4)$.
\label{sim:randomconsistent}}
\begin{tabular}{ccrcccc}
\hline
     &  &  \multicolumn{2}{c}{Estimator  $\bar{Z}$ } & \multicolumn{2}{c}{Estimator $Z_1$}\\
 $N$ & $T$  &  AbsRelBias & SdAbsRelBias & AbsRelBias & SdAbsRelBias \\ 
  \hline    
10 & 30 & 0.0315 & 0.3649 & 0.1351 & 0.6581 \\ 
  & 100 & 0.0333 & 0.1949 & 0.0382 & 0.5996 \\ 
     \cline{2-6}
  100 & 30 & 0.0047 & 0.0503 & 0.0193 & 0.1704 \\ 
  & 100 & 0.0057 & 0.0347 & 0.0054 & 0.1653 \\ 
     \cline{2-6}
  1000 & 30 & 0.0020 & 0.0122 & 0.0013 & 0.0514 \\ 
   & 100 & 0.0011 & 0.0070 & 0.0029 & 0.0481 \\ 
 \hline 
\end{tabular}
\end{center}
\end{table}

\begin{table}
\begin{center}
\captionof{table}{Absolute relative bias and standard deviations of the proposed estimators $Z_1$ and $\bar{Z}$ for Model 3 when edge persistence probabilities are modeled as $p_{ij} = p_{i}p_{j}$ with $p_{i}$ drawn from $\text{Beta}(1,4)$, for $i=1,\cdots,N$.
\label{sim:productrandomconsistent}}
\begin{tabular}{ccrcccc}
\hline
     &  &  \multicolumn{2}{c}{Estimator  $\bar{Z}$ } & \multicolumn{2}{c}{Estimator $Z_1$}\\
 $N$ & $T$  &  AbsRelBias & SdAbsRelBias & AbsRelBias & SdAbsRelBias \\ 
  \hline    
10 & 30 & 0.3208 & 0.5165 & 0.4897 & 0.8497 \\ 
 & 100 & 0.1125 & 0.4485 & 0.4202 & 0.9954 \\ 
  \cline{2-6}
  100 & 30 & 0.0028 & 0.0987 & 0.0811 & 0.3657 \\ 
   & 100 & 0.0134 & 0.0668 & 0.0849 & 0.3683 \\ 
    \cline{2-6}
  1000 & 30 & 0.0025 & 0.0250 & 0.0013 & 0.1029 \\ 
   & 100 & 0.0020 & 0.0140 & 0.0190 & 0.1067 \\ 
 \hline 
\end{tabular}
\end{center}
\end{table}

\section{Reproductive Number for the SIR Process}

While the TCM can be leveraged for any spreading process over a network, here we examine the spread of an infectious disease. To model disease spread, we employ compartmental models, a class of models that divides the population into different groups with respect to disease state. These models assign transition rules, allowing individuals to move between different states. Here, we use the susceptible-infectious-recovered (SIR) model, which assumes that individuals obtain perfect immunity once they recover from the disease. Under this model, individual nodes can be in any of three states: susceptible, infectious, and recovered. In the susceptible state, the node has not been infected but could become infected if it came into contact with an infectious node. That is, the node has no immunity to the disease. In the infected state, the node is infectious and can infect others it comes into contact with. Finally, in the recovered state, the node has recovered from the illness and can no longer infect others or be reinfected. With a stochastic model, nodes move through states probabilistically. Due to the discrete nature of our data, we utilize a discrete-time approach wherein events are defined by transition probabilities per unit time, as opposed to the transition rates used in a continuous time framework. A contact between an infectious and a susceptible node will result in a transmission with probability $\beta$. Similarly, at any given time step a node will recover with probability $\gamma$. For simplicity, we assume that at each time step, each edge has persistence probability $p$. 

Let $R_0$ denote the average number of transmissions from the initially infected node and $R_*$ denote the average number of transmissions in the early stages of the epidemic excluding those from the initial infection. For both a static and dynamic network, $R_0 = \tau \sum_k kp_k$, where $\tau$ is the transmission probability for a link between an infected node and a susceptible node and $p_k$ is the probability a node has degree $k$ in the configuration model. For a static network,$R_* = \tau \sum_k q_k(k-1)$, where $q_k$ is the excess degree distribution of a node with degree $k$, i.e., $q
_k = kp_k/\sum_k kp_k$. \cite{volz2009} 

Before further discussion of the reproductive number, we recall an important tool in studying disease spread on a network: the probability generating function (PGF). Suppose the PGF for the degree distribution of the initial configuration model network is $g(x) = \sum_{k} p_k x^k$, where $p_k$ is the probability that a randomly chosen edge has degree $k$. The PGF of the excess degree distribution is $g_1(x) = \sum_{k} q_{k} x^{k-1} $, where $q_{k} = kp_{k}/\sum_{k} k p_k$. Thus, we have the following relations: 
\begin{align}
g'(1) &= \sum_k kp_k\\
g''(1) &= \sum_k k(k-1) p_k\\
    g'_1(1) &= \sum_{k} (k-1) q_{k} = \sum_{k} (k-1) kp_k / \sum_{k} k p_k = g''(1)/g'(1).
\end{align} 

We now examine the reproductive number at the early stage of the SIR spreading process on the temporal configuration model. 
\begin{proposition}
\label{proR}
     The reproductive number at the early stage of a discrete time SIR spreading process on the temporal network is given by $R_* =\tau [ (1-\gamma)(1-p)/\gamma +  (1 - p + \gamma p ) g''(1) /\big( \gamma g'(1)  \big)] $, where $\tau = \frac{\beta }{1- p(1-\beta)(1- \gamma)}.$
\end{proposition}
\textit{Proof:}

Following the idea of Volz and Meyers  \cite{volz2009}, we derive the reproductive number for the process through four steps as follows: 
\begin{enumerate}
\item Find the probability of transmission $\tau$ for each connection.
\item Find the probability generating function (PGF) of the number of contacts $M$ for a node with degree k. 
\item Find the PGF $\tilde{H}_1(x)$ of the number of contacts of the selected node proportional to the concurrent degree, but subtracting one transitory contact. 
\item Find the PGF $H_1(x)$ of the number of transmissions caused by the selected node proportional to the concurrent degree, but subtracting one transitory contact. 
\end{enumerate}
Then the reproductive number $R_*$ at the early stage of the epidemic is the derivative of $H_1(x)$ evaluated at $x =1$.

We first find the probability of transmission $\tau$ for each connection. Notice that for a given edge, the time $X$ until the edge has broken follows a geometric distribution, where $X \sim $ \text{Geo}$(1-p)$. Similarly, the time $Y$ until an infected node transmits to a susceptible neighbor satisfies $Y \sim $ \text{Geo}$(\beta)$, and the time $Z$ until an infected node recovers satisfies $Z \sim $ \text{Geo}$(\gamma)$. For an infected-susceptible connection, the overlap time $U$ while the infected node is still infectious and the edge is still connected is $U \sim \text{min}(X,Z)$. Notice that $X$ and $Z$ are independent, therefore $U \sim \text{Geo}(1- p (1-\gamma))$ by the properties of a geometric distribution.

So, the probability that the infected node transmits the disease to the susceptible in at most $k$ steps is $P(Y \leq  k-1) = 1 - (1-\beta)^k$. Therefore, the transmission probability $\tau$ during the overlap time is 
\begin{align}
\label{transmission}
\tau &= E\left(1- (1-\beta)^U\right)= \sum_{k\geq 1} P(U=k)\left[1 - (1-\beta)^k\right] = \sum_{k \geq 1} \left[1 - (1-\beta)^k\right] [p(1- \gamma)]^{k-1} [1-p(1-\gamma)] \notag \\
 & = [1-p(1-\gamma)] \sum_{k \geq 1} [p(1- \gamma)]^{k-1}  - [1-p(1-\gamma)](1-\beta) \sum_{k \geq 1} [p(1-\beta)(1- \gamma)]^{k-1} \notag \\
 &= [1-p(1-\gamma)] \frac{1}{1 - p(1- \gamma)} - [1-p(1-\gamma)](1-\beta) \frac{1}{1- p(1-\beta)(1- \gamma)} \notag\\
 &= 1 - \frac{[1-p(1-\gamma)](1-\beta)}{1- p(1-\beta)(1- \gamma)}= \frac{\beta }{1- p(1-\beta)(1- \gamma)}.
\end{align}


We then find the PGF of number of contacts $M$ for an infected node with degree $k$ during time period $\ell$. For simplicity, we first construct the PGF for $M$ corresponding to nodes with degree $1$. For a node of degree $1$, the number of edges swapped during $\ell$ time steps follows Binomial$(\ell,(1-p))$. Therefore, the probability that the number of contacts for a node of degree $1$ is equal to $m$ during the period of $\ell$ steps is $ \binom{\ell}{ m-1} (1-p)^{m-1}p^{\ell-m+1}$. Thus, for a node with degree $1$, the PGF of the number of contacts $M$ is 
$ \sum_{m\geq 1} \binom{\ell}{ m-1} (1-p)^{m-1}p^{\ell-m+1} x^m = x  \sum_{m\geq 1} \binom{\ell}{ m-1} [x(1-p)]^{m-1}p^{\ell-m+1} = x [x(1-p) + p]^\ell.$ Therefore, for a node with degree $k$, the PGF for the number of contacts $M$ is $\{x [x(1-p) + p]^\ell\}^k$.

Next, we provide the PGF  $\tilde{H}_1(x)$ of the number of transitory contacts of a selected node proportional to the concurrent degree, but subtracting one transitory contact. Construction of $\tilde{H}_1(x)$ includes the following three steps: 
\begin{enumerate}
\item[(a)] Sum over the infectious period $\ell$.
\item[(b)] Sum over all PGFs of all possible degree $k$ while weighting for the excess degree, $q_k = k p_k/(\sum k p_k)$.
\item[(c)] Add an extra term, denoted as $f_\ell(x)$, corresponding to the PGF that accounts for the concurrent partnership of the infected node being considered, but that excludes the contact that infected the node being considered. 
\end{enumerate}
Following the same argument as for the PGF of number of contacts for a node with degree $1$, we have $f_\ell(x) = \sum_{m\geq 1} \binom{\ell}{ m-1} (1-p)^{m-1}p^{\ell-m+1} x^{m-1} = \sum_{m\geq 1} \binom{\ell}{ m-1} [x(1-p)]^{m-1}p^{\ell-m+1} = [x(1-p) + p]^\ell$. Notice that the probability of the infectious period $\ell$ is equal to the probability that the infected node recovers at time $\ell+1$. Therefore, we have 

\begin{align*}
   \tilde{H}_1(x) & = \sum_{\ell \geq 0} Pr(Z = \ell+1) \left[f_\ell(x) + \sum q_k (x f_\ell(x))^{k-1}\right] \\
   & = \sum_{\ell \geq 0} (1-\gamma)^\ell \gamma \left\{ \big(x(1-p) + p\big)^\ell + \sum_{k\geq 1} q_k \left[x \big(x(1-p) + p\big)^\ell\right]^{k-1}\right\}\\ 
& = \sum_{\ell \geq 0} (1-\gamma)^\ell \gamma \big( x(1-p) + p \big)^\ell  + \sum_{\ell \geq 0} (1-\gamma)^\ell \gamma \sum_{k\geq 1} q_k \left[x \big(x(1-p) + p\big)^\ell\right]^{k-1} \\ 
& = \gamma \sum_{\ell \geq 0} [(1-\gamma)\big(x(1-p) + p\big)]^\ell  + \gamma  \sum_{k\geq 1} q_k x^{k-1} \sum_{\ell \geq 0} \left[ (1-\gamma) \left( x(1-p) + p\right)^{k-1}\right]^\ell \\
   & =\gamma /[1- (1-\gamma) \big( x(1-p) + p\big) ]+  \gamma \sum_{k\geq 1} q_k x^{k-1}/\left[1 -   (1-\gamma) \big( x(1-p) + p\big)^{k-1}\right]
\end{align*}
where the summation exists as long as $x < \big(1/(1-\gamma\big)^{1/(k-1)}-p)/(1-p)$ for all $k \geq 2$. Therefore,
\begin{align}
\label{Avecontact0}
  \tilde{H}'_1(1) & = \gamma (1-\gamma)(1-p)/\gamma^2 + \gamma \sum_{k\geq 1} \big( \gamma q_k (k-1) + q_k  (k-1) (1-\gamma) (1-p) \big)/\gamma^2 \notag \\
  & =   (1-\gamma)(1-p)/\gamma + (1 - p + \gamma p )  \sum_{k\geq 1} q_k (k-1)/\gamma,   
\end{align}

or 
\begin{equation}
\label{Avecontact}
    \tilde{H}'_1(1) =   (1-\gamma)(1-p)/\gamma +  (1 - p + \gamma p ) g''(1) /\big( \gamma g'(1)  \big).
\end{equation}

Finally, we find the PGF $H_1(x)$ of the number of transmissions caused by the selected node proportional to the concurrent degree, but subtracting one transitory contact. The PGF of number of transmissions corresponding to $k$ transitory contacts is $(1- \tau + \tau  x )^k$. Since the PGF of the number of transmissions is the summation of all possible transitory contacts, 
\begin{equation}
H_1(x) = \sum P(\#\text{transitory} = k) (1- \tau + \tau  x )^k  = \tilde{H}_1(1- \tau + \tau  x ).
\end{equation}

Then we have, 
\begin{equation}
    R_* = (dH_1(x)/dx)_{x=1} = (d \tilde{H}_1(1- \tau + \tau  x )/dx)_{x=1} = \tau \tilde{H}'(1).
\end{equation}

Applying (\ref{transmission}) and (\ref{Avecontact}), the reproductive number $R_*$ is

\begin{equation}
R_* =\tau [ (1-\gamma)(1-p)/\gamma +  (1 - p + \gamma p ) g''(1) /\big( \gamma g'(1)  \big)], 
\end{equation}

where $\tau = \frac{\beta }{1- p(1-\beta)(1- \gamma)}$. Proposition \ref{proR} is thus proved. $\hfill  \square$

Next, we examine these results under specific degree distributions. 
\begin{itemize}
    \item When the degree distribution follows a Poisson distribution with mean $\lambda$, $p_k = \lambda^k \exp(-k)/k!$, its corresponding PGF is $g(x) = \exp\big(\lambda(x-1)\big)$. Then $g'(1) = \lambda$ and $g''(1) = \lambda^2$. Therefore, $R_* = \tau [ (1-\gamma)(1-p)/\gamma +  (1 - p + \gamma p ) \lambda / \gamma  ] $. For $p= 1$, this gives us $R_* = \tau \lambda$, which agrees with the results derived in  \cite{volz2009}.

\item When the degree distribution satisfies $p_k = 1$, if $k = M$ for some $ 1 \leq M \leq N$ where $N$ is the network size, then the PGF is given by $g(x) = x^M$. So $g''(1)/g'(1) = M-1$, and therefore $R_* = \tau [ (1-\gamma)(1-p)/\gamma +  (1 - p + \gamma p ) (M-1) / \gamma   ] $. Notice that if $k=N$, we have a fully connected network. Under this scenario, the reproductive number is given by $R_* = \tau [ (1-\gamma)(1-p)/\gamma +  (1 - p + \gamma p ) (N-1) / \gamma$]. For $p=1$, this reduces to $R_* = \tau (N-1)$. 
\end{itemize}

Additionally, when the temporal network is static over time, $p = 1$. The reproductive number at the early stages is then $R_* = \tau g''(1)/g'(1)  $, which agrees with the $R_*$ value for a static network introduced above. When the temporal network evolves as independent draws from a configuration model, then $p=0$. Under this scenario, the reproductive number at the early stage is $R_* = \tau  [ (1-\gamma)/\gamma +   g''(1) /\big( \gamma g'(1)  \big)] = \beta  [ (1-\gamma)/\gamma +   g''(1) /\big( \gamma g'(1)  \big)] $.

\section{Data Analysis}

\begin{figure}
  \centering
  \makebox[0pt]{\includegraphics[width=\textwidth]{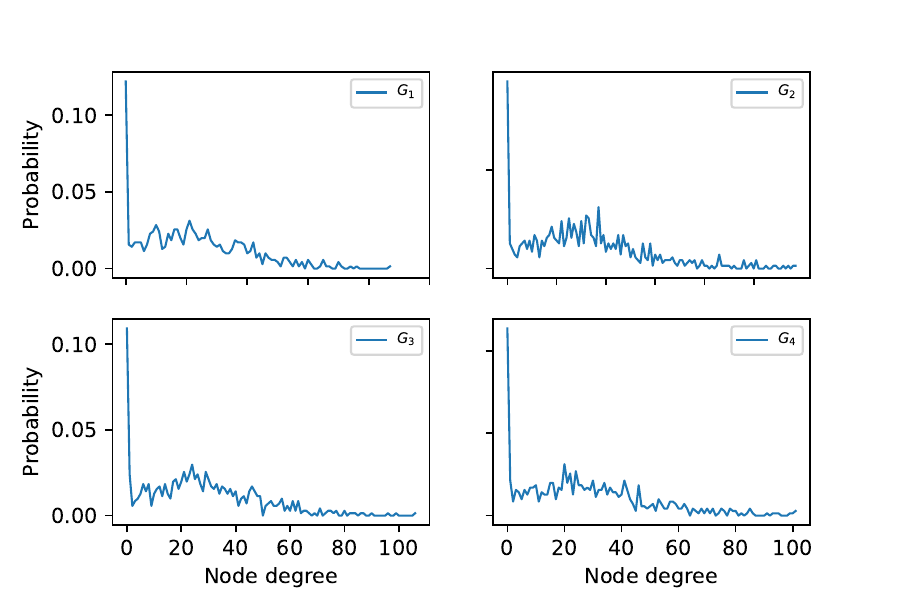}}
  \caption{\textbf{Degree distributions of weekly empirical networks $G_1, G_2, G_3, G_4$}.}
  \label{fig:Degree distribution}
\end{figure}

In this section, we show how to use the three proposed TCMs to fit empirical network data and examine the fit of the generated temporal networks. The empirical data in this study was collected by the Copenhagen Network Study data and is publicly available.  \cite{sapiezynski2019interaction} This data set contains information about the connectivity patterns of 706 students at the Technical University of Denmark over 28 days in February 2014. During the study period, participants agreed to use loaner cell phones from researchers as their primary phones. The proximity patterns were collected using Bluetooth where the approximate pairwise distances of phones were obtained using the received signal strength indicator (RSSI) every five minutes. We assigned a connection between two persons if there was at least one strong Bluetooth ping with RSSI $\geq \text{-75dBm}$  \cite{hambridge2021}. Because the empirical data reflects student proximity patterns, they fluctuate significantly on a daily basis. Students were more likely to connect during the week if they were in the same class, but on weekends, their proximity patterns were likely driven by their personal contact networks. Our models cannot capture the weekday-weekend variability because they require a comparable number of edges at each time instance. Therefore, we further processed the daily network data by combining each of the seven daily networks into a single weekly network. In particular, we established a weekly network by taking the union of the daily networks of each weekly batch of 7 daily networks. As a result, the period of 28 days yielded 4 weekly networks, denoted as $\{ G_1, G_2, G_3, G_4\}$. Figure \ref{fig:Degree distribution} shows their degree distributions.

 To fit the TCM models, we used $G_1$ as the starting point and counted the number of edges in $G_1$ that were retained in $G_2$ and $G_3$. For Model 1, we calculated the fixed edge persistence rate as the ratio of the number of edges in $G_1$ that persisted into $G_2$. For Models 2 and 3, we assumed that the edge-level and node-level persistence rates were generated from Beta distributions. The first and second moments of these Beta distributions were estimated by the ratios of the number of edges in $G_1$ that persisted into $G_2$ and $G_3$, respectively. Using the estimated first two moments, we determined the shape and scale parameters of the Beta distributions. The empirical weekly data gave us the estimate $\hat{p} = 0.476$ for Model 1; the Beta distributions for Model 2 and Model 3 were estimated as Beta$(0.975, 1.074)$ and Beta$(1.171, 1.562)$, respectively. With the estimated model parameters, we used $G_1$ as the initial network and used the TCM models to obtain predicted networks $\hat{G}_2^{(M_k)}, \hat{G}_3^{(M_k)},\hat{G}_4^{(M_k)}$ for the empirical networks $G_2, G_3, G_4$, respectively. Here $k = 1, 2, 3$ represents Model 1, Model 2, and Model 3, respectively. In addition, as a benchmark, we employed the naive temporal configuration model, in which all edges were broken and rewired randomly at each time step. We refer to this configuration model as Model 0. 
 
 To assess the performance of the TCM models, we compared the distances between the degree distributions of networks generated with Model 0, Model 1, Model 2, and Model 3 with the empirical networks $G_2, G_3, G_4$. We used the total variance distance, $D = \frac{1}{2} \sum_{i=1}^k \vert p_i - q_i \vert$, and the Hellinger distance, $H = \frac{1}{\sqrt{2}} \sqrt{\sum_{i=1}^k (\sqrt{p_i} - \sqrt{q_i})^2} $, where $P = (p_1,\cdots,p_k)$ and $Q = (q_1,\cdots,q_k)$ are discrete distributions. We denote the distances between the predicted networks $\hat{G}_2^{(M_k)}$, $\hat{G}_3^{(M_k)}$, $\hat{G}_4^{(M_k)}$ and the corresponding empirical networks $G_2, G_3, G_4$ as $D_2^{(M_k)}, D_3^{(M_k)}, D_4^{(M_k)}$ and $H_2^{(M_k)}, H_3^{(M_k)}, H_4^{(M_k)}$, respectively. We then computed the average distances $\bar{D}^{(M_k)} = \frac{1}{3}\left(D_2^{(M_k)} +  D_3^{(M_k)} + D_4^{(M_k)}\right)$ and $\bar{H}^{(M_k)} = \frac{1}{3} \left(H_2^{(M_k)} +  H_3^{(M_k)} + H_4^{(M_k)}\right)$ to assess the performance of the models. To make a fair assessment, we ran the simulation 100 times, calculated the means and standard deviations of $\bar{D}^{(M_k)}$ and $\bar{H}^{(M_k)}$ based on these 100 runs, and used them as final goodness-of-fit metrics to assess the performance of the models. In particular, for run $i$, we calculated the average distances $\bar{D}^{(M_k,i)}$ and $\bar{H}^{(M_k,i)}$; the final metrics for comparing the four models are the means and standard deviations of these metrics. 
 
 Table \ref{realdata:distance} demonstrates that Model 2 and Model 3 yield the smallest distances and therefore fit the empirical data best. Model 1 performed worse than Models 2 and 3, and Model 0 had the worst performance. The results demonstrate that model fit generally improves from Model 1 to Model 2 to Model 3. The improvement in model fit was to be expected because Model 0 uses data about the initial network only; Model 1 requires additional knowledge of the number of edges that remain in the initial network $G_1$ after one time step, and Models 2 and 3 require knowledge of the number of edges that remain after one and two time steps. In this example, which should be regarded as being primarily for illustrative purposes, Models 2 and 3 have similar performance. In general, we anticipate Model 3 to provide a potentially better fit in settings where the behavior of individuals is driven less by external factors, such as class schedule here, and instead depends more on the individual choices of the actors.

\begin{table}
\begin{center}
\captionof{table}{Assessment of model goodness-of-fit with means and standard deviations of total variation distances and Hellinger distances over 100 runs.
\label{realdata:distance}}
\begin{tabular}{ccrcc}
\hline
Model   &  \multicolumn{1}{c}{Total variation distance} & \multicolumn{1}{c}{Hellinger distance}\\
  
  \hline    
Model 0  & 0.2400 $ \pm$ 0.0058  & 0.2803 $\pm$ 0.0046 \\   

Model 1 & 0.2338 $ \pm$ 0.0057 & 0.2727 $\pm$ 0.0050\\ 

Model 2 & 0.2336 $ \pm$ 0.0055   & 0.2718 $ \pm$ 0.0045 \\ 

Model 3 & 0.2330  $ \pm$ 0.0054  & 0.2706 $\pm$ 0.0054 \\  
 \hline 
\end{tabular}
\end{center}
\end{table}

\section{Discussion and Conclusion}
We introduced the Temporal Configuration Model (TCM), a family of generative models for a temporal network, as well as approaches for estimating model parameters. The proposed generative model is simple and flexible. The modeling framework allows the edge persistence probability to be fixed, generated from a distribution, or constructed from the corresponding node-level persistence probabilities. We demonstrated how to use the modeling framework by applying it to data from the Copenhagen Network Study with promising results. Therefore, the generative model can be adapted to fit a variety of real-world settings. 

We proposed consistent estimators for the model parameters and provided the convergence rate of the proposed estimators. We found that only using information from the first one or two time steps can produce a good estimator for model parameters. We additionally showed that, when the persistence probability is constant across edges, using the network evolution process can give us a more precise estimator with a faster convergence rate. However, when persistence probabilities are generated at random or are the product of the node-level persistence probabilities, care must be taken when using an estimator that relies on averages over the entire course of the network's evolution. Instead, it might be advantageous to consider alternative modeling strategies, such as those in which persistence probabilities are periodically updated. If additional information on this time interval is available, then a design that incorporates the network evolution process will give us a finer estimator with an accelerated convergence rate. 

In addition, we investigated an SIR spreading model over the temporal network under the scenario where edge persistence probability is constant. We provided an explicit formula for the epidemic's reproductive number at the early stage of the pandemic for this modeling scenario.

Further research directions might involve examining a good estimator for the periodic information in Model 2 and Model 3, extending the temporal model to include network size growth information, and investigating various epidemic characteristics of spreading processes on the temporal network of the three models. All of these potential developments, as well as the broad range of applications for temporal networks, offer an exciting future for the study of time-varying networks. \cite{holme15reviewTCM,  gage2020reviewTCM, mehdi23reviewTCM}
With technological advances, particularly Bluetooth technology, network data is becoming more accessible than ever. How we can appropriately employ these new datasets for public health insights is an interesting subject that requires further research. We believe that statistical techniques with robust theoretical underpinnings are the foundation for leveraging the vast amounts of data available for the common good.

\paragraph{Acknowledgement} 
The authors thank Dr. Marc Lipsitch and Dr. Jeff Miller at Harvard Chan School of Public Health for their insightful feedback.

\paragraph{Contributions} 
J.P.0., T.M.L., and H.H. designed the research; T.M.L. and H.H. performed the research; T.M.L., H.H., and J.P.O. wrote and edited the paper. J.P.O supervised the research. T.M.L. and H.H. are shared first co-authors.

\paragraph{Change of Institution} 
T.M.L. started the project at Harvard Chan School of Public Health, Harvard University, Boston, Massachusetts, U.S.A.  

\paragraph{Funding Statement}
H.H. was supported by a Harvard University Department of Biostatistics scholarship and a U.S. Government scholarship. T.M.L and J.P.O. were supported by a National Institutes of Health award (NIAID R01 AI138901). The funding sources had no role in study design, data analysis, data interpretation, or the writing of the paper.

\paragraph{Data and materials availability}
The empirical network data is publicly available at The Copenhagen Networks Study interaction data. \cite{sapiezynski2019interaction} Python code used in this study is publicly available at https://github.com/onnela-lab/temporal-configuration-model.

\paragraph{Competing Interests}
None.

\bibliographystyle{unsrt}

\begin{thebibliography}{1}

\bibitem{newman2018networks}
\textsc{ Newman, Mark} (2018).
\newblock Networks.
\newblock \textit{Oxford University Press}.



\bibitem{roland20review}
\textsc{ Roland, Molontay and Marcell, Nagy} (2020).
\newblock Two Decades of Network Science as seen through the co-authorship network of network scientists
\newblock \textit{ArXiv}.


\bibitem{delva2016connectdot}
\textsc{ Wim, Delvaa, Gabriel, E. Leventhal and St{\'e}phane Helleringer} (2016).
\newblock Connecting the dots: network data and models
in HIV epidemiology
\newblock \textit{AIDS}.

\bibitem{brea2018ego}
\textsc{ Brea, L. Perry, Bernice, A. Pescosolido and Stephen, P. Borgatti} (2018).
\newblock Egocentric Network Analysis.
\newblock \textit{Cambridge University Press}.


\bibitem{holme2012temporal}
\textsc{ Holme, Petter and Saram{\"a}ki, Jari} (2012).
\newblock Temporal networks.
\newblock \textit{Physics reports}.

\bibitem{holme15reviewTCM}
\textsc{ Holme, Peter} (2015).
\newblock Modern temporal network theory: A colloquium.
\newblock \textit{ArXiv}.

\bibitem{holme21maptemporal}
\textsc{ Holme, Peter and Jari, Saram\"{a}ki } (2021).
\newblock A map of approaches to temporal networks.
\newblock \textit{ArXiv}.


\bibitem{holme13optimalstatic}
\textsc{ Holme, Peter} (2013).
\newblock Epidemiologically optimal static networks from
temporal network data.
\newblock \textit{PLoS Comput. Biol.}.

\bibitem{gage2020reviewTCM}
\textsc{ Gage, Jordan, Samuel, Winer and Taban, Salem} (2020).
\newblock The current status of temporal network analysis
for clinical science: Considerations as the
paradigm shifts?
\newblock \textit{Journal of Clinical Psychology}.


\bibitem{mehdi23reviewTCM}
\textsc{ Mohammad, Mehdi Hosseinzadeh, Mario, Cannataro, Pietro, Hiram Guzzi and Riccardo, Dondi} (2023).
\newblock Temporal networks in biology and medicine: a survey on models,
algorithms, and tools.
\newblock \textit{Network Modeling Analysis in Health Informatics and Bioinformatics}.

\bibitem{hambridge2021}
\textsc{ Hambridge, Hali L, Kahn, Rebecca and Onnela, Jukka-Pekka} (2021).
\newblock Examining sars-cov-2 interventions in residential colleges using an empirical network.
\newblock \textit{International Journal of Infectious Diseases}.

\bibitem{wu22temporalcovid}
\textsc{ Mincheng, Wu, Chao, Li, Zhangchong, Shen and others} (2022).
\newblock Use of temporal contact graphs to understand
the evolution of COVID-19 through contact
tracing data.
\newblock \textit{Communication Physics}.

\bibitem{vanhems2013estimating}
\textsc{ Vanhems, Philippe, Barrat, Alain, Cattuto, Ciro and others} (2013).
\newblock Estimating potential infection transmission routes in hospital wards using wearable proximity sensors.
\newblock \textit{PloS one}.

\bibitem{sapiezynski2019interaction}
\textsc{ Sapiezynski, Piotr, Stopczynski, Arkadiusz, Lassen, David Dreyer and Lehmann, Sune} (2019).
\newblock Interaction data from the copenhagen networks study.
\newblock \textit{Scientific Data}.

\bibitem{neto2021combining}
\textsc{ Neto, Onicio Leal, Haenni, Simon, Phuka, John and others} (2021).
\newblock Combining wearable devices and mobile surveys to study child and youth development in Malawi: implementation study of a multimodal approach.
\newblock \textit{JMIR Public Health and Surveillance}.


\bibitem{perra12activitydriventemporal}
\textsc{ N. Perra, B. Gonçalves, R. Pastor-Satorras and A. Vespignani} (2012).
\newblock Activity driven modeling of time varying
networks.
\newblock \textit{Scientific Reports}.

\bibitem{vestergaard14linknodetemporal}
\textsc{ Vestergaard, Christian L., G{\'e}nois, Mathieu and Barrat,  Alain} (2014).
\newblock How memory generates heterogeneous dynamics in temporal networks.
\newblock \textit{Physical Review}.



\bibitem{tiago2017modelcommunity}
\textsc{ Tiago, P. Peixoto and Martin, Rosvall} (2017).
\newblock Modelling sequences and temporal networks with
dynamic community structures.
\newblock \textit{Nature Communication}.






\bibitem{zhang2017randomdynamic}
\textsc{ Xiao, Zhang and Cristopher, Moore and M. E. J. Newman} (2017).
\newblock Random graph models for dynamic networks.
\newblock \textit{The European Physical Journal B}.






\bibitem{Bailey18CM}
\textsc{ Bailey, K. Fosdick and Daniel, B. Larremore and 
Joel, Nishimura and 
Johan, Ugander} (2018).
\newblock Configuring random graph
models with fixed degree sequences.
\newblock \textit{SIAM Review}.








\bibitem{Alex2022}
\textsc{ Alexander, D. and Davy, P.} (2022).
\newblock On the consistency of incomplete U-statistics under infinite
second-order moments.
\newblock \textit{Statistics and Probability Letters}.




\bibitem{volz2009}
\textsc{ Volz, E. and Meyers, L.A.} (2009).
\newblock Epidemic thresholds in dynamic
contact networks.
\newblock \textit{Journal of the Royal Society Interface}.















\end{thebibliography}

\end{document}